\documentclass[10pt,times]{iopart}
\usepackage{epsfig}
\usepackage{graphics}
\usepackage{graphicx}
\usepackage{bm}
\usepackage{iopams}
\usepackage{color}

\begin{document}


\newcommand{\ajith}[1]{\textcolor{magenta}{\textit{Ajith: #1}}}
\newcommand{\yanbei}[1]{\textcolor{red}{\textit{Yanbei: #1}}}
\newcommand{\sascha}[1]{\textcolor{blue}{\textit{Sascha: #1}}}
\newcommand{\alicia}[1]{\textcolor{green}{\textit{Alicia: #1}}}
\definecolor{purple}{rgb}{0.5,0,0.5}
\newcommand{\jtw}[1]{\textcolor{purple}{\textit{JTW: #1}}}
\newcommand{\lr}[1]{\textcolor{red}{\textit{LR: #1}}}

\newcommand{\be}{\begin{equation}}
\newcommand{\ee}{\end{equation}}
\newcommand{\ber}{\begin{eqnarray}}
\newcommand{\eer}{\end{eqnarray}}
\newcommand{\ie}{i.e.}
\newcommand{\dt}{{\rm d}t}
\newcommand{\dtheta}{{\rm d}\theta}
\newcommand{\diota}{{\rm d}\iota}
\newcommand{\dphi}{{\rm d}\phi}

\newcommand{\rhat}{\hat{r}}
\newcommand{\iotahat}{\hat{\iota}}
\newcommand{\phihat}{\hat{\phi}}
\newcommand{\hc}{\textsf{h}}

\title[Phenomenological templates]{Phenomenological template family
for black-hole coalescence waveforms}


\author{P.~Ajith$^1$, S.~Babak$^2$, Y.~Chen$^2$, M.~Hewitson$^1$, B.~Krishnan$^2$,
J.~T.~Whelan$^2$, B.~Br\"ugmann$^3$, P.~Diener$^{4,5}$,
J.~Gonzalez$^3$, M.~Hannam$^3$, S.~Husa$^3$,
M.~Koppitz$^2$, D.~Pollney$^2$, L.~Rezzolla$^{2,5}$,
L.~Santamar\'ia$^3$, A.~M.~Sintes$^{2,6}$, U.~Sperhake$^3$, and
J.~Thornburg$^{2,7}$}

\address{$^1$~Max-Planck-Institut f\"ur Gravitationsphysik 
(Albert-Einstein-Institut) and Leibniz Universit\"at Hannover, 
Callinstr.~38, 30167~Hannover, Germany}

\address{$^2$~Max-Planck-Institut f\"ur Gravitationsphysik 
(Albert-Einstein-Institut), 
Am~M\"uhlenberg 1, 14476~Golm, Germany}

\address{$^3$~Theoretisch-Physikalisches Institut, Friedrich Schiller Universit\"at Jena,
Max-Wien-Platz 1, 07743~Jena, Germany}

\address{$^4$~Center for Computation \& Technology, Louisiana State University, \\
Baton Rouge, LA, USA}

\address{$^5$~Department of Physics and Astronomy, Louisiana State University, \\
Baton Rouge, LA, USA}

\address{$^6$~Departament de F\'{\i}sica, Universitat de les Illes
Balears, Cra. Valldemossa Km. 7.5, E-07122 Palma de Mallorca, Spain}

\address{$^7$~School of Mathematics, University of Southampton,
Southampton SO17~1BJ, England}


\begin{abstract}

Recent progress in numerical relativity has enabled us to model the
non-perturbative merger phase of the binary black-hole coalescence
problem.  Based on these results, we propose a phenomenological family
of waveforms which can model the inspiral, merger, and ring-down
stages of black hole coalescence.  We also construct a template bank
using this family of waveforms and discuss its implementation in the
search for signatures of gravitational waves produced by black-hole
coalescences in the data of ground-based interferometers. This
template bank might enable us to extend the present inspiral searches
to higher-mass binary black-hole systems, \textit{i.e.}, systems with
total mass greater than about 80 solar masses, thereby increasing the
reach of the current generation of ground-based detectors.

\end{abstract}

\section{Introduction}
\label{sec:intro}

The first generation of ground-based gravitational wave detectors
~\cite{LIGOStatus:2006,GEOStatus:2006,VirgoStatus:2006} are currently
operating at unprecedented levels of sensitivity and the LIGO
detectors, in particular, have attained their design goals over a
broad frequency range. The data from these detectors has been used to
search for a wide variety of gravitational-wave sources including
coalescing binary black-hole systems (see \textit{e.g.},
\cite{S2-inspiral, S3S4-inspiral}).  In parallel with these
experimental and observational achievements, a series of breakthroughs
has occurred in numerical simulations of binary black hole systems
\cite{Pretorius:2005gq,Campanelli:2005dd,Baker05a}.  Long-term
evolutions of inspiralling black holes that last for several orbits
have been obtained with several independent
codes~\cite{Pretorius:2006tp,Campanelli:2006gf,Baker:2006yw,Sperhake:2006cy,Bruegmann:2006at,Scheel-etal-2006:dual-frame,Herrmann:2007ac,Koppitz:2007ev},
and accurate gravitational-wave signals have been computed.  It is now
possible, in principle, to use these numerical-relativity results in
astrophysical searches for gravitational waves.  However, the high
computational cost of these simulations makes it unfeasible to
numerically generate all the necessary waveforms to cover the
parameter space that needs to be searched.  It is therefore necessary,
at the present time, to use results from post-Newtonian (PN) theory to
extend the waveforms obtained from numerical relativity (NR).  The
issue of matching NR and PN waveforms for equal-mass binary black-hole
systems, and using numerical relativity results in gravitational-wave
searches has been considered previously
\cite{Buonanno:2006ui,Baumgarte:2006en,Baker:2006ha,Pan:2007nw}.  We
generalize this to unequal mass systems and suggest a phenomenological
template bank parametrized only by the masses of the two individual
black holes.  This template bank could be used to search for binary black-hole
signals in data from current and future generations of
gravitational-wave detectors. Our phenomenological approach is
motivated by the work of Buonanno \etal \cite{BCV1}, but using only
\emph{physical} parameters and generalized to include recent results from numerical
relativity.  

In this paper we combine \emph{restricted} 3.5PN waveforms
\cite{3.5PNBlachetEtal:2004} with results from NR simulations to
construct \textit{``hybrid''} waveforms for the quasi-circular
inspiral of non-spinning binaries with possibly unequal masses.
Restricting ourselves to the leading-order quadrupole modes, we find
that the hybrid waveforms can be approximated by phenomenological
analytical waveforms with fitting factors $\geq 0.99$ in the total
mass range between 30 and 130 $M_\odot$ for Initial LIGO
\cite{Abramovici92}. For our analysis we use numerical waveforms obtained with two
independent codes: \textit{(i)} long waveforms (12 cycles) from equal
mass binaries, provided by the AEI-CCT groups, using their CCATIE code
~\cite{Alcubierre:2002kk} based on the Cactus framework~\cite{Cactus}
and Carpet mesh-refinement driver~\cite{Schnetter:2003rb};
\textit{(ii)} waveforms from an unequal mass parameter study presented
by the Jena group in \cite{Gonzales06tr}, which have been obtained
with the BAM code \cite{Bruegmann:2006at}. An analysis of these
waveforms focusing on ring-down and higher modes has been presented in
\cite{Berti:2007fi}.
     
The rest of this paper is organized as follows:
Section~\ref{sec:numrelintro} summarizes the numerical simulations and
how numerical waveforms have been computed.  In
Section~\ref{sec:Matching} hybrid waveforms are produced by matching PN
and NR waveforms. In Section~\ref{sec:phenomtemplates}, we propose a
family of phenomenological waveforms in the Fourier domain, and study
their impact for detection and parameter estimation by computing the
fitting factors of the phenomenological waveforms with the hybrid
ones. We also parametrize the best-matched phenomenological
waveforms in terms of the physical parameters.
Section~\ref{sec:range} shows the astrophysical range of a search
using the full coalescence waveforms and a preliminary comparison with other searches.
Finally, Section~\ref{sec:conc} concludes with a summary of our
results and plans for future work.


\section{Numerical simulations}
\label{sec:numrelintro}

Both the BAM~\cite{Bruegmann:2006at} and CCATIE~\cite{Alcubierre:2002kk} codes are
finite-difference mesh-refinement codes solving the Einstein equations
within the ``moving puncture''
framework~\cite{Campanelli:2005dd,Baker05a,Hannam:2006vv,Koppitz:2007ev}.

In the wave-zone, sufficiently far away from the source, the spacetime
metric can be accurately described as a perturbation of a flat
background metric; let $h_{ab}$ denote the metric perturbation where
$a,b$ denote spacetime indices.  Let $t$ be the time coordinate used
in the numerical simulation to foliate the spacetime by spatial
slices.  Working, as usual, in the transverse-traceless (TT) gauge,
all the information about the radiative degrees of freedom is
contained in the spatial part $h_{ij}$ of $h_{ab}$, where $i,j$ denote
spatial indices.  Let us use a coordinate system $(x,y,z)$ on a
spatial slice so that the $z$-axis is along the total angular momentum
of the binary system at the starting time.  Let $\iota$ be the
inclination angle from the $z$-axis, and let $\phi$ be the phase angle
and $r$ the radial distance coordinates so that $(r,\iota,\phi)$ are
standard spherical coordinates in the wave-zone.  

Working in the TT gauge, the radiative degrees
of freedom in $h_{ab}$ can be written, as usual, in terms of two
polarizations $h_+$ and $ h_\times$:
\begin{equation}
  \label{eq:2}
  h_{ij}  = h_+ (\mathbf{e}_{+})_{ij} +
  h_\times(\mathbf{e}_\times)_{ij},
\end{equation}
where $\mathbf{e}_{+,\times}$ are the usual basis tensors for TT
tensors in the wave frame
\begin{equation}
  \label{eq:9}
  (\mathbf{e}_+)_{ij} = \iotahat_i \iotahat_j -
  \phihat_i \phihat_j\,, \qquad \textrm{and} \qquad 
  (\mathbf{e}_\times)_{ij} = \iotahat_{i}\phihat_{j} + \iotahat_{j}\phihat_{i}\,.
\end{equation}
Here $\iotahat$ and $\phihat$ are the unit vectors in the $\iota$ and
$\phi$ directions respectively. The wave, of course, propagates in the
radial direction.

In our numerical simulations, the gravitational waves are extracted by
two distinct methods. The first one uses the Newman-Penrose Weyl
tensor component $\Psi_4$ (see e.g. \cite{Stewart}) which, in an
appropriate gauge is a measure of the outgoing transverse
gravitational radiation in an asymptotically flat spacetime. By
measuring that the peeling property (whereby $\Psi_4$ falls off
as $1/r$) is satisfied, we have determined
that the gauge we are using does provide a good approximation to
within the accuracy required for this study. In the
wavezone it can be written in terms of the complex strain $\hc = h_+ -
\mathrm{i} h_\times$ as~\cite{Teukolsky:1973ha},
\begin{equation}
  \hc = \lim_{r\rightarrow\infty} \int^t_0 \dt^\prime
  \int^{t^\prime}_0 \dt^{\prime\prime} \, \Psi_4.  \label{eq:psi4}
\end{equation}
An alternative method for wave extraction, which has a long history in
numerical relativity, determines the waveform via gauge-invariant
perturbations of a background Schwarzschild spacetime, via the
Zerilli-Moncrief formalism (see~\cite{Nagar:2005ea} for a review). In
terms of the even ($Q^{(+)}_{\ell m}$) and odd ($Q^{(\times)}_{\ell
  m}$) parity master functions, the gravitational wave strain
amplitude is then given by
\begin{equation}
 \hc = \frac{1}{\sqrt{2}r}
  \sum_{\ell,m}\left(Q^+_{\ell m} - \mathrm{i}
    \int^t_{-\infty}Q^\times_{\ell m}(t^\prime)\,\dt^\prime\right)
    Y^{-2}_{\ell m} + \mathcal{O}\left(\frac{1}{r^2}\right).
  \label{eq:zerilli}
\end{equation}

Results from the BAM code have used the Weyl tensor component $\Psi_4$
and Eq. (\ref{eq:psi4}), with the implementation described
in~\cite{Bruegmann:2006at}. While the CCATIE code computes waveforms
adopting both methods, the AEI-CCT waveforms used here were computed
using the perturbative extraction and Eq. (\ref{eq:zerilli}).  Beyond
an appropriate extraction radius, the two methods for determining $\hc$
are found to agree very well for moving-puncture black-hole evolutions
of the type considered here~\cite{Koppitz:2007ev}.

It is useful to discuss gravitational radiation fields in terms of
spin-weighted $s=-2$ spherical harmonics $Y^{s}_{\ell m}$, and in this
paper we will only consider the dominant $\ell=2,\ m=\pm 2$ modes (see
\cite{Berti:2007fi} for the higher $\ell$ contribution in the unequal
mass case), with basis functions
\begin{equation}
  Y^{-2}_{2-2} \equiv \sqrt{\frac{5}{64\pi}} \left(1 -\cos \iota \right)^2
       e^{-2 \mathrm{i}\phi}, \quad
  Y^{-2}_{22} \equiv \sqrt{\frac{5}{64\pi}} \left( 1 +\cos \iota \right)^2
       e^{2 \mathrm{i}\phi}.
\end{equation}
Our ``input'' NR waveforms correspond to the projections
\begin{equation}
\hc_{\ell m} \equiv \langle Y^{-2}_{\ell m}, \hc \rangle = 
	\int_0^{2\pi} \dphi \int_0^{\pi}
      \hc \, \overline{Y^{-2}_{\ell m}}\, \sin \iota\,\diota\,
      \label{eq: scalar_product},
\end{equation}
of the complex strain $\hc$, where the bar denotes complex
conjugation.  In the cases considered here, we have equatorial
symmetry so that $\hc_{22} = \overline{ \hc_{2-2} }$, and
\begin{equation}
  \hc(t) = \sqrt{\frac{5}{64\pi}} e^{2 \mathrm{i}\phi} \left(
    \left(1 + \cos \iota \right)^2       \hc_{22}(t)
    + \left(1 - \cos \iota \right)^2  \bar \hc_{22}(t) \right). 
\end{equation}
In practice, we choose $\iota=0$, thus $\hc(t) = 4
\sqrt{\frac{5}{64\pi}} \, \hc_{22}(t) \approx 0.631\, \hc_{22}(t)$.



\section{Matching post-Newtonian and numerical relativity waveforms}
\label{sec:Matching}

\begin{figure}
  \begin{center}
    \includegraphics[width=12cm]{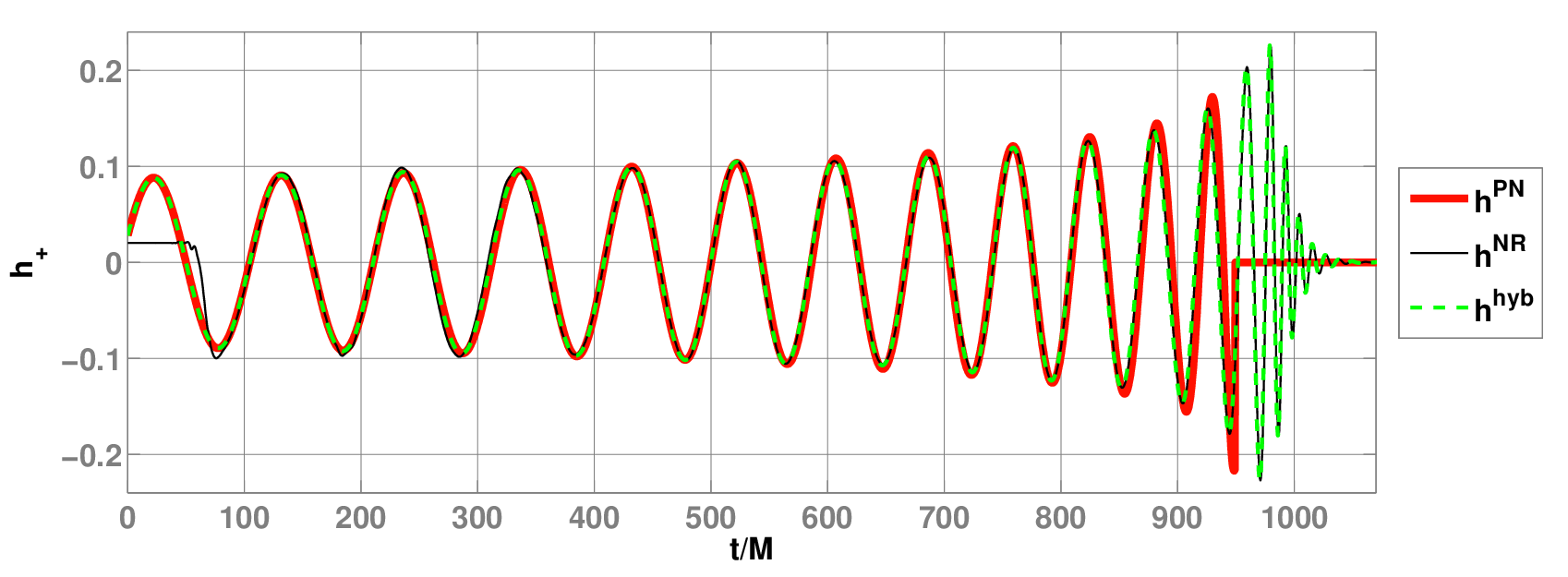}
    \caption{NR waveform (black) from an equal-mass simulation, along with the `best-matched' 
    3.5PN waveform (red). The Hybrid waveform constructed from the above is also shown (dashed line).}
    \label{fig:MatchTimeDomWaves}
  \end{center}
\end{figure}

Once the PN and NR waveforms are generated, we produce a set of hybrid
waveforms by matching them in an overlapping time interval $t_1 \leq t <
t_2$. The obvious assumption in this procedure is that such an
overlapping region exists and that in it both approaches yield the
correct waveforms.

Each time-domain waveform $h(t,{\bm \mu})$ is parametrized by a vector
${\bm \mu} = \{M, \eta, \phi_0, t_0\}$, where $M \equiv m_1+m_2$ is
the total-mass of the binary, $\eta \equiv {m_1 m_2}/{(m_1+m_2)^2}$ is
the symmetric mass-ratio, $\phi_0$ is the initial phase and $t_0$ is
the start time of the waveform.  We match the PN waveforms $h^{\rm
PN}_{+,\times}(t,{\bm \mu})$ and the NR waveforms $h^{\rm
NR}_{+,\times}(t,{\bm \nu})$
\footnote{The parameters ${\bm \nu}$ are taken from the same set as
${\bm \mu}$, but with different values.} by minimizing the square
difference between the respective polarizations. \textit{i.e.},
\ber
\delta &\equiv& {\rm min}_{{\bm \mu}, a} \left[ \sum_{i=+,\times} \, \int_{t_1}^{t_2}  
\left[h_i^{^{\rm PN}}(t,{\bm \mu})-a\, h_i^{^{\rm NR}}(t,{\bm \nu})\right]^2 \, \dt \right].
\eer
The minimization is carried over the parameters ${\bm \mu}$ of the PN
waveform and an amplitude scaling factor $a$~\footnote{The amplitude
scaling factor was introduced in order to accommodate possible errors
in the amplitude of the NR waveforms due to, for example, the finite
radius of the extraction sphere. The best-matched value of $a$,
however, was found to be $1 \pm 0.08$.}.
The hybrid waveforms are produced by combining the `best-matched' PN
waveforms and the NR waveforms in the following way:
\ber
h^{\rm hyb}_{+,\times}(t, {\bm \nu}) \equiv \left\{ \begin{array}{ll}
h^{^{\rm PN}}_{+,\times}(t,{\bm \mu_0}) & \textrm{if $t < t_1 $}\\ \\
a_0 \, \tau \, h^{^{\rm NR}}_{+,\times}(t,{\bm \nu}) 
+ (1 - \tau)\, h^{^{\rm PN}}_{+,\times}(t,{\bm \mu_0}) & \textrm{if $t_1
  \leq t < t_2 $}\\ \\
a_0 \, h^{^{\rm NR}}_{+,\times}(t,{\bm \nu}) & \textrm{if $t_2 \leq t$}
\end{array} \right.
\label{eq:HybWave}
\eer
where $\bm \mu_0$ and $a_0$ denote the values of $\bm \mu$ and $a$ for
which $\delta$ is minimum, and $\tau = (t-t_1)/(t_2-t_1)$ is a
linearly-increasing weighting function, such that $0 \leq \tau < 1$.

An example set of hybrid waveforms is shown in
Fig.~\ref{fig:MatchTimeDomWaves}.  The numerical waveform (black line)
from an equal-mass ($\eta = 0.25$) simulation by the AEI-CCT group is
matched with a 3.5PN inspiral waveform (red line) over the matching
region $100 M \leq t < 850M$.  The hybrid waveform (dashed line) is
constructed by combining the above as per Eq.(\ref{eq:HybWave}).

The robustness of the matching procedure can be tested by computing
the overlaps between hybrid waveforms constructed with different
matching regions. If the overlaps are very high, this can be taken as
an indication of the robustness of the matching procedure. A more
detailed discussion of this will be presented in~\cite{NRDApaper2}.

\begin{figure}
  \begin{center}
    \includegraphics[height=6cm]{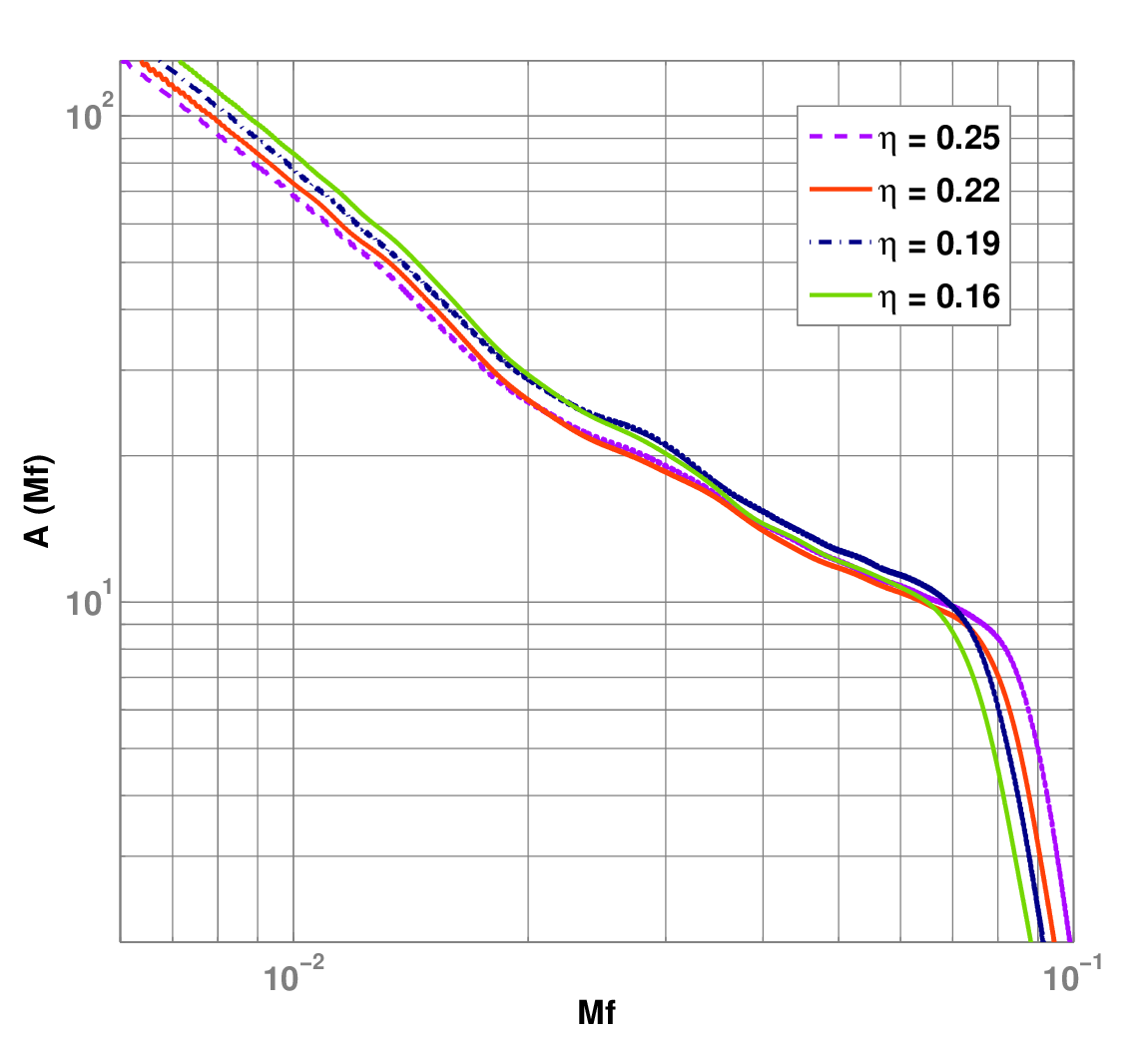}
    \includegraphics[height=6cm]{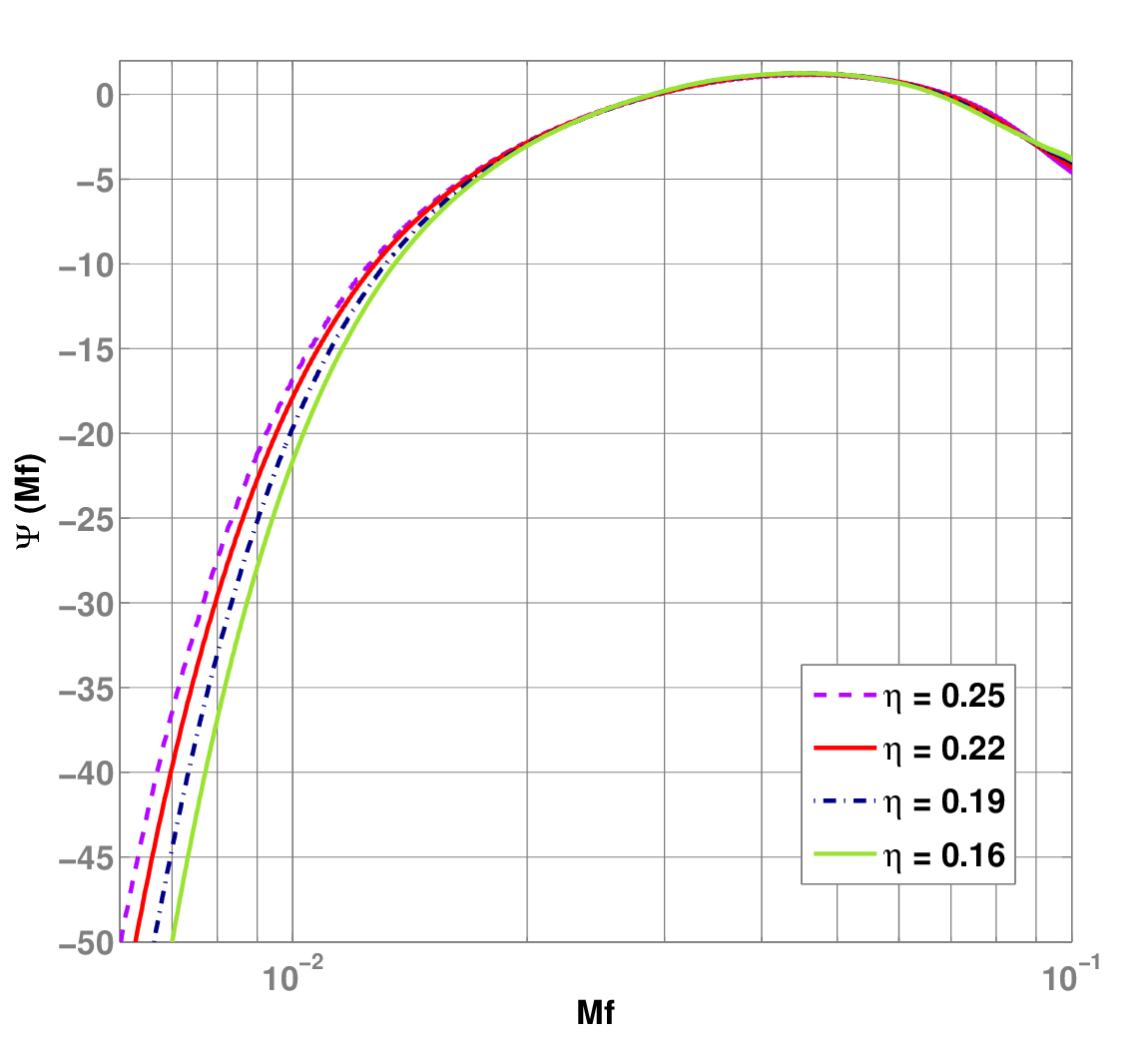}
    \caption{Fourier domain magnitude (left) and phase (right) of the
    hybrid waveforms. Symmetric mass-ratio $\eta$ of each waveform is
    shown in the legends.}
    \label{fig:TargetWaveFreqDomain}
  \end{center}
\end{figure}

Fig.~\ref{fig:TargetWaveFreqDomain} shows the hybrid waveforms of
different mass-ratios in the Fourier domain. In particular, the
panel on the left shows the amplitude of the waveforms in the Fourier
domain, while the panel on the right shows the phase. These waveforms
are constructed by matching 3.5PN waveforms with the NR waveforms from
the unequal mass ($0.16 \leq \eta \leq 0.25$) simulations by the Jena
group. In the next section, we try to parametrize these Fourier domain
waveforms in terms of a set of phenomenological parameters.

\section{The phenomenological template bank}
\label{sec:phenomtemplates}

An obvious issue when using the above constructed hybrid waveforms
directly as detection templates is that it might be computationally
very expensive to compute enough NR waveforms to cover the entire
parameter space densely enough. In this section, we propose a
phenomenological waveform family which has more than 99 \% overlaps
with the hybrid waveforms in the detection band of the Initial LIGO
detectors. We also show how this phenomenological waveform family can
be mapped to the physical parameters ($M$ and $\eta$), so that the
template bank, at the end, is two-dimensional.

\begin{figure}
  \begin{center}
    \includegraphics[width=8.5cm]{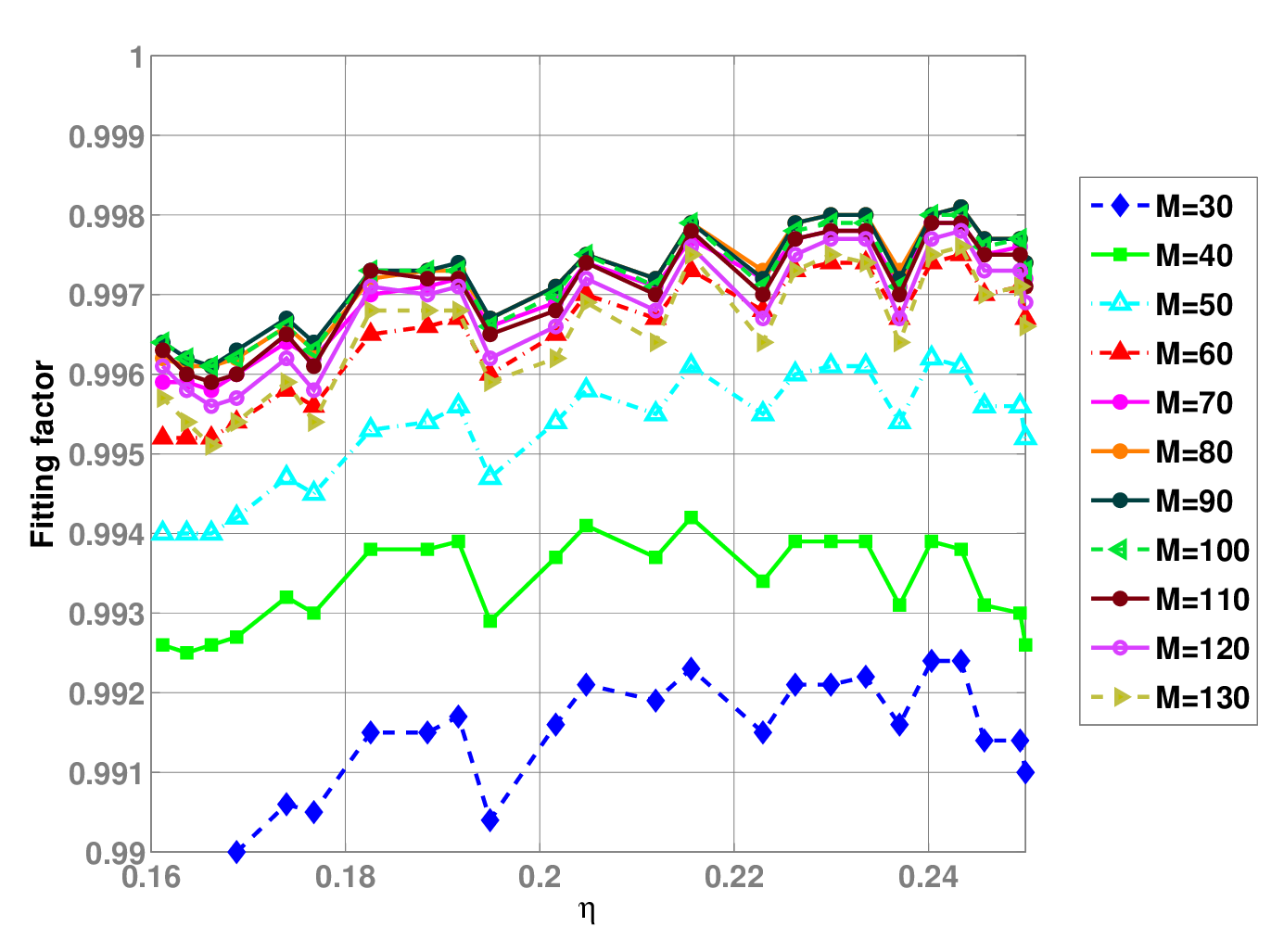}
    \caption{Fitting factors of the hybrid waveforms with the
    phenomenological waveform family.  Horizontal axis shows the
    symmetric mass ratio of the binary, while different
    colours/markers correspond to different total masses. }
    \label{fig:FitFactorsJenaUM}
  \end{center}
\end{figure}

\subsection{Phenomenological waveforms}
We write our phenomenological waveform in the Fourier domain as 
\be
u(f) \equiv {\cal A}_{\rm eff}(f) \, 
	{\mathrm e}^{{\mathrm i}\Psi_{\rm eff}(f)}.  
\ee 
where ${\cal A}_{\rm eff}(f)$ is the amplitude of the waveform in the
frequency domain, which we choose to write in terms of a set of
\textit{``amplitude parameters''} $\bm \alpha = \{f_{\rm merg}, f_{\rm
ring}, f_{\rm cut}, \sigma\}$ as
\be
{\cal A_{\rm eff}}(f) \equiv 
\left\{ \begin{array}{ll}
\left(f/f_{\rm merg}\right)^{-7/6} & \textrm{if $f < f_{\rm merg}$}\\\\
\left(f/f_{\rm merg}\right)^{-2/3} & \textrm{if $f_{\rm merg} \leq f < f_{\rm ring}$}\\\\
w \, {\cal L}(f,f_{\rm ring},\sigma) & \textrm{if $f_{\rm ring} \leq f < f_{\rm cut}$}
\end{array} \right. 
\ee
In the above expression, 
\begin{equation}
{\cal L}(f,f_{\rm ring},\sigma ) \equiv \left(\frac{1}{2 \pi}\right)
\frac{\sigma}{(f-f_{\rm ring})^2+\sigma^2/4} \,,
\end{equation}
represents a Lorentzian function of width $\sigma$ centered around
$f_{\rm ring}$. The normalisation constant $w$ is chosen in such a way
that ${\cal A}_{\rm eff}(f)$ is continuous across the ``transition''
frequency $f_{\rm ring}$, \textit{i.e.},
\be
w \equiv \frac{\pi \sigma}{2} \left(\frac{f_{\rm ring}}{f_{\rm merg}}\right)^{-2/3}\,,
\ee
where we use $f_{\rm cut}$ as the cutoff frequency of the template and
$f_{\rm merg}$ as the frequency at which the power-law changes from
$f^{-7/6}$ to $f^{-2/3}$ (as noted previously in
\cite{Buonanno:2006ui} for the equal-mass case).

Taking our motivation from the stationary-phase expansion of the
gravitational-wave phase, we write the effective phase $\Psi_{\rm
eff}(f)$ as an expansion in powers of $f$.
\begin{equation}
\fl\qquad \Psi_{\rm eff}(f)  =  2 \pi f t_0 + \phi_0 + \psi_0 \, f^{-5/3}
+ \psi_2 \,f^{-1} +\psi_{3} \,f^{-2/3} + \psi_{4} \,f^{-1/3} +
\psi_{6} \,f^{1/3} \,,
\end{equation}
where $t_0$ is the time of arrival, $\phi_0$ is the frequency-domain
phase offset, and $\bm \beta =
\{\psi_0,\,\psi_2,\,\psi_3,\,\psi_4,\,\psi_6\}$ are the
\textit{``phase parameters''}, that is the set of phenomenological
parameters describing the phase of the waveform.

\begin{figure}[tbh]
  \begin{center}
    \includegraphics[width=11cm]{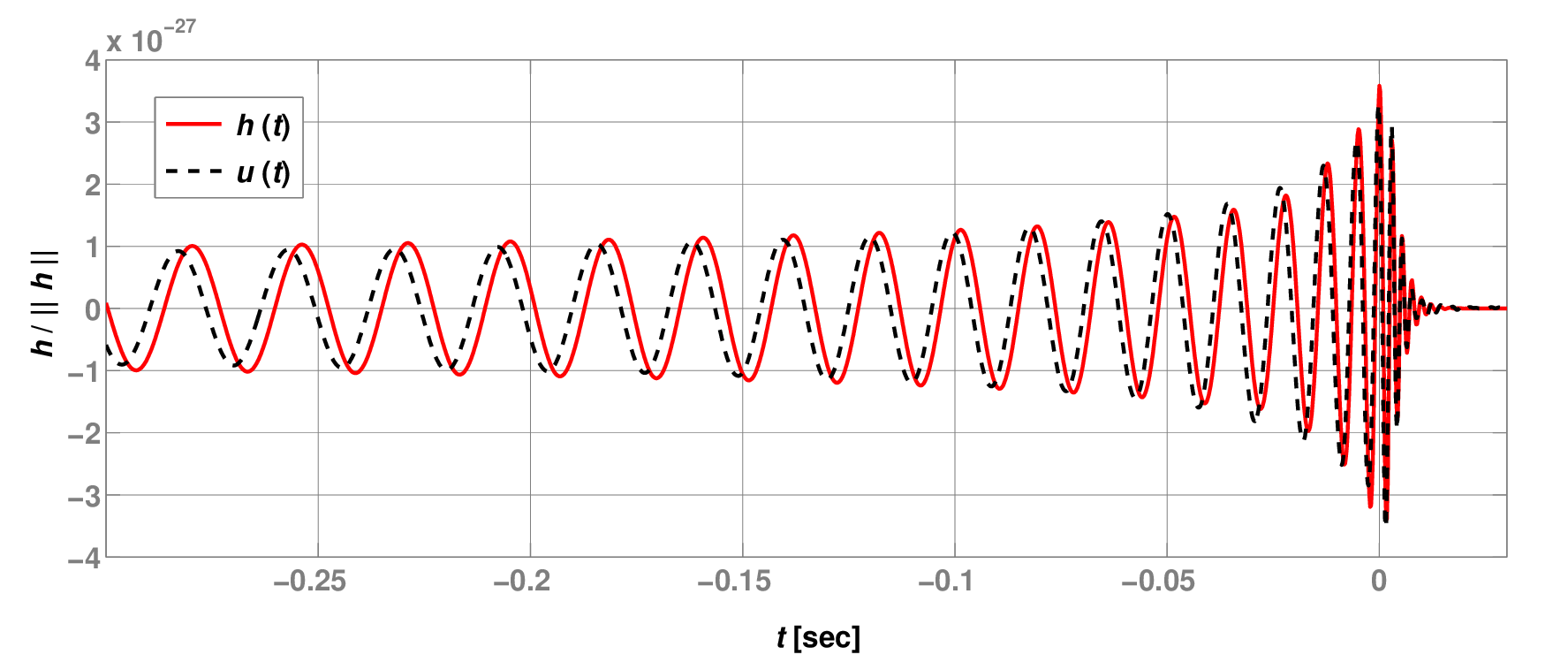}
    \caption{Hybrid waveform $h(t)$ and the best-matched
      phenomenological waveform $u(t)$ in the time domain for a $M=
      40\, M_\odot, \eta = 0.25$ binary system. $u(t)$ is computed by
      taking the inverse Fourier transform of the phenomenological
      waveform $u(f)$. Both waveforms are normalised with respect to
      the Initial LIGO noise spectrum.}
    \label{fig:TimeDomHybAndPhenWave}
  \end{center}
\end{figure}

The fitting factors~\cite{Apostolatos:1995} of the hybrid waveforms
with the family of phenomenological waveforms are shown in
Fig.~\ref{fig:FitFactorsJenaUM} over the parameter range of $30 \leq
M/M_\odot \leq 130$ and $0.16 \leq \eta \leq 0.25$, using the Initial
LIGO noise spectrum. It is quite apparent that the fitting factors are
almost always greater than 0.99, thus underlining the effectiveness of
the phenomenological waveforms in reproducing the hybrid ones. As an example, 
in Fig.~\ref{fig:TimeDomHybAndPhenWave}, we plot the hybrid waveform
and the best-matched phenomenological waveform from the $M= 40\, M_\odot, \eta = 0.25$ binary. 

\subsection{From phenomenological to physical parameters}

\begin{figure}
  \begin{center}
    \includegraphics[width=12.5cm]{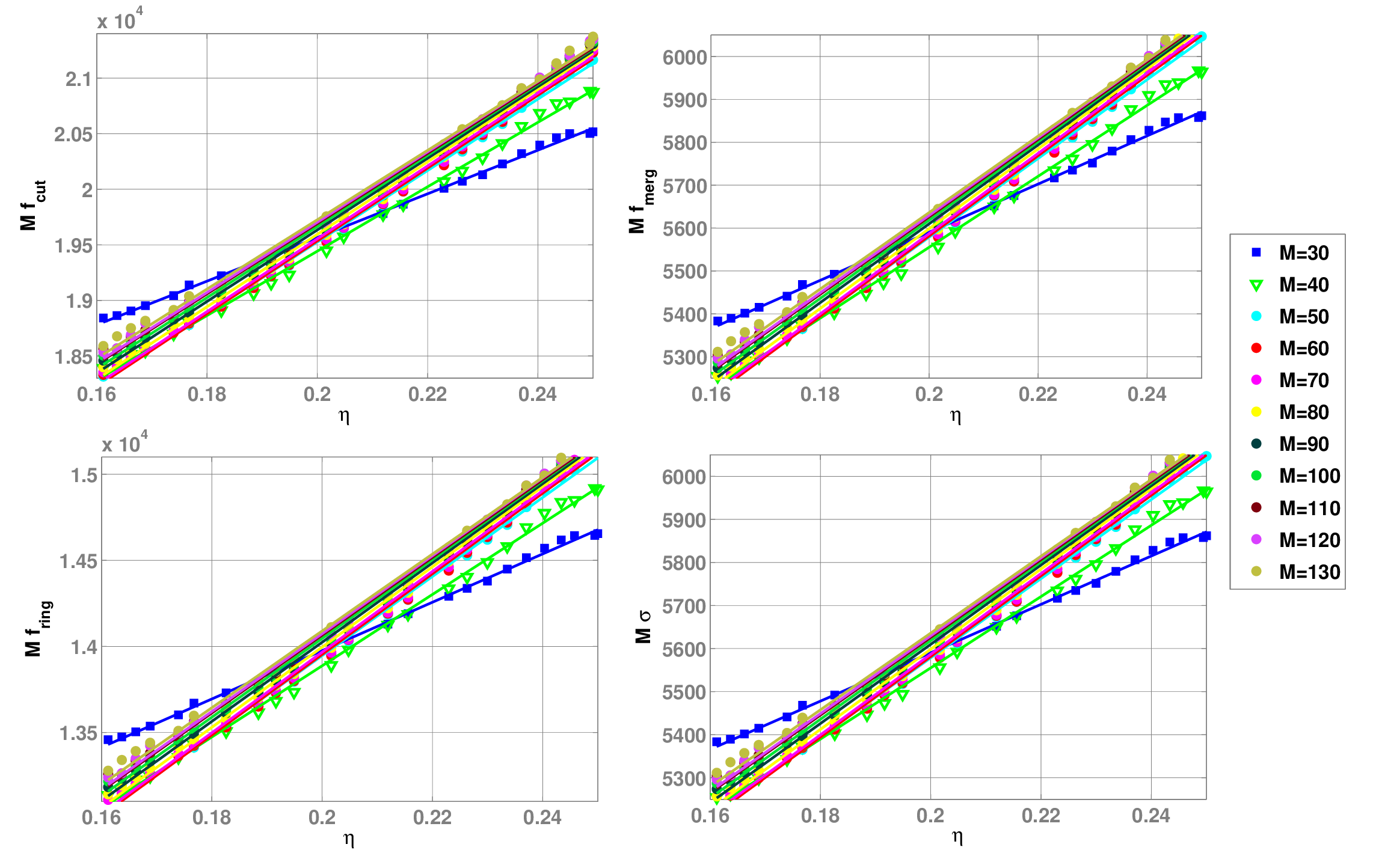}
    \caption{Best-matched amplitude parameters $\bm{\alpha}_{\rm max}$
    in terms of the physical parameters of the binary. The horizontal
    axis shows the symmetric mass-ratio of the binary and different
    colors/markers correspond to different total masses. Linear
    polynomial fits to the data points are also shown.}
    \label{fig:BestMatchedAmpParams}
  \end{center}
\end{figure}
\begin{figure}
  \begin{center}
    \includegraphics[width=12.5cm]{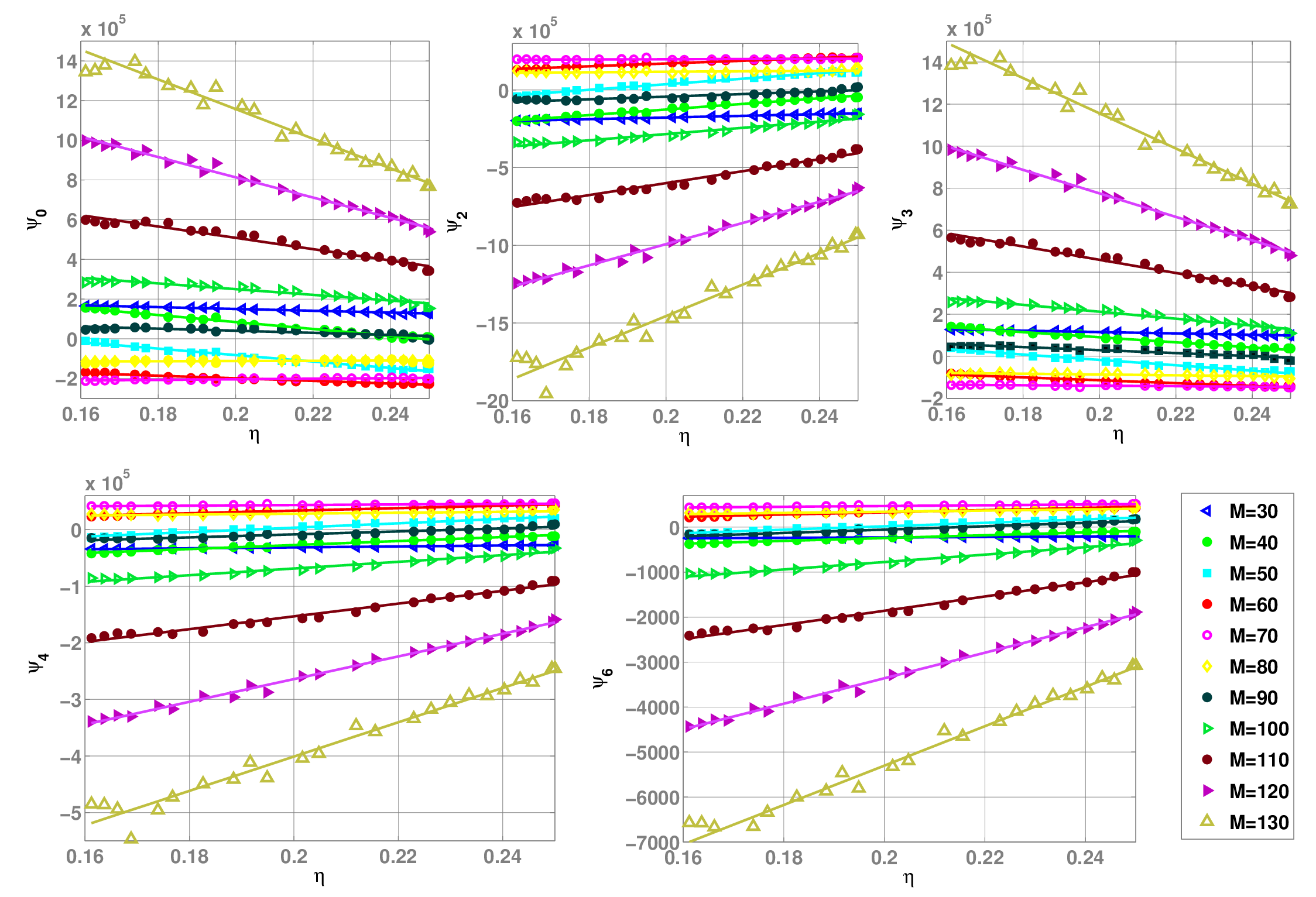}
    \caption{The same as in Fig.~\ref{fig:BestMatchedAmpParams} but
    for the phase parameters $\bm{\beta}_{\rm max}$.}
    \label{fig:BestMatchedPhaseParams}
  \end{center}
\end{figure}


It is possible to reparametrise the best-matched phenomenological
waveforms in terms of the physical parameters of the hybrid
waveforms. In Fig.~\ref{fig:BestMatchedAmpParams}, we plot the
amplitude parameters ${\bm \alpha}_{\rm max}$ of the best-matched
phenomenological waveforms against the physical parameters of the
binary.  Similarly, the phase parameters ${\bm \beta}_{\rm max}$ of
the best-matched phenomenological waveforms are plotted against the
physical parameters of the binary in
Fig.~\ref{fig:BestMatchedPhaseParams}.  The linear polynomial fits
to the data points serve as a numerical ``look-up table'' to go from the
physical parameters $\{M,\eta\}$ to the best-matched phenomenological
parameters $\{\bm{\alpha}_{\rm max},\bm{\beta}_{\rm max}\}$. Thus the
template bank lives on a two-dimensional manifold (parametrised by $M$
and $\eta$) embedded in a higher-dimensional space.

It might be worth stressing that the search will be carried out over
$M$ and $\eta$, and \emph{not} over the phenomenological
parameters. The phenomenological parameters ${\bm \alpha}$ and ${\bm
\beta}$, are constrained by the numerical look-up tables (see
Figs.~\ref{fig:BestMatchedAmpParams} and
\ref{fig:BestMatchedPhaseParams}), and only serve as an intermediate
step in generating the template waveforms.



\section{The astrophysical range and comparison with other searches}
\label{sec:range}

\begin{figure}
  \begin{center}
    \includegraphics[height=7cm]{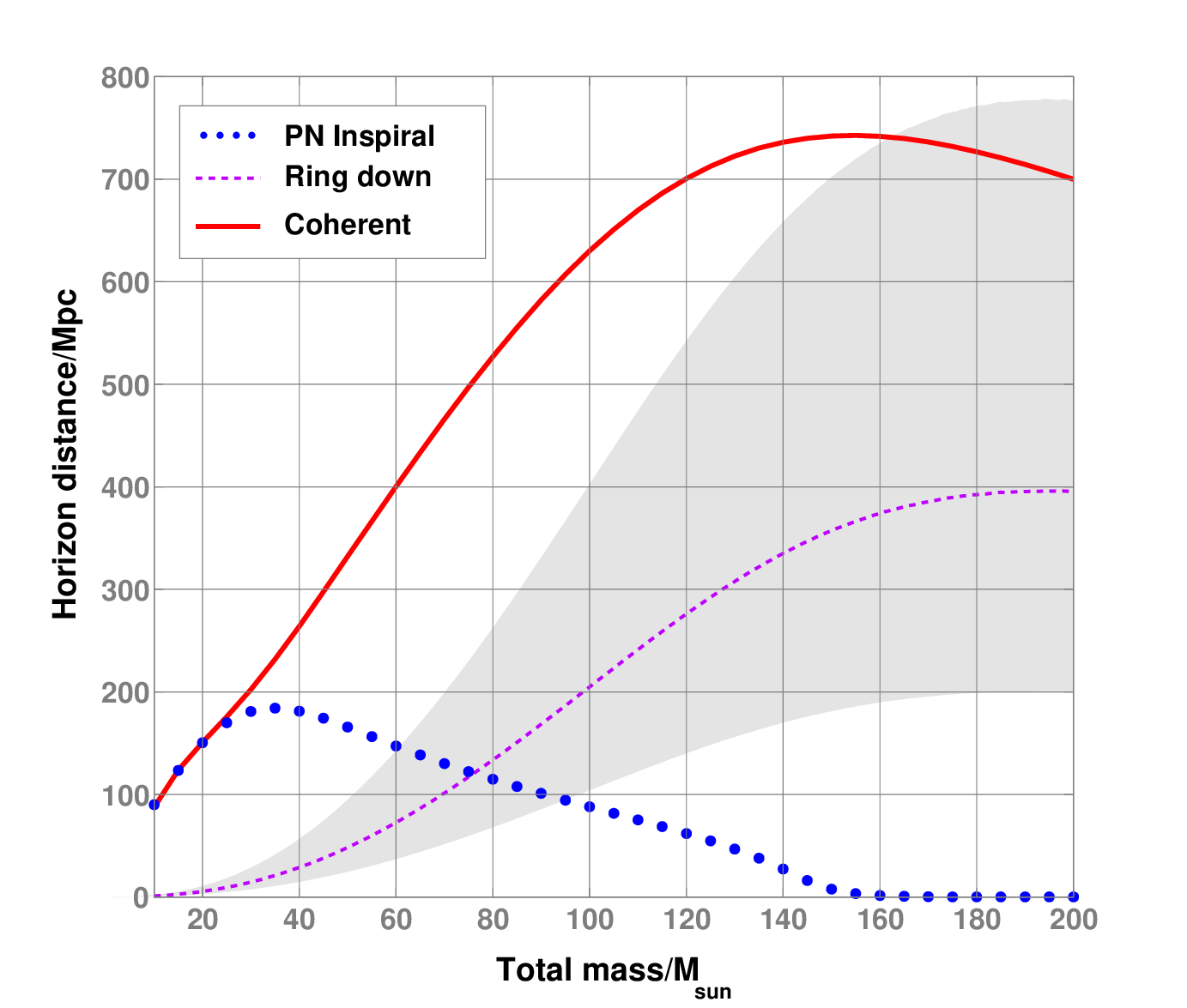}
    \caption{A preliminary assessment of the phenomenological template
      bank. This figure plots, as a function of the total mass, the
      distance to an optimally-oriented, equal-mass binary which can
      produce an optimal SNR of 8 at the Initial LIGO noise spectrum.
      The dotted line corresponds to a search using PN templates
      truncated at ISCO, the dashed line corresponds to a search using
      ring-down templates, and the solid line to a search using the
      template family proposed in this paper. The ring down horizon
      distance is computed assuming that $\epsilon=0.7\%$ of the black
      hole mass is radiated in the ring down stage. Since the value of
      this parameter has some amount of uncertainty in it, we have also included the shaded
      region in the plot corresponding to $0.18\% \leq \epsilon \leq
      2.7\%$.}
     \label{fig:HorDist}
  \end{center}
\end{figure}

The template family proposed in this paper can be used for coherently
searching for all the three stages (inspiral, merger, and ring-down)
of the binary coalescence, thus making this potentially more sensitive
than searches which look at the three stages separately.
Fig.~\ref{fig:HorDist} compares the sensitivity of the searches using
different template families. What is plotted here is the distance at
which an optimally-oriented, equal-mass binary would produce an
optimal signal-to-noise ratio (SNR) of 8 at the Initial LIGO noise
spectrum. The dotted line corresponds to a search using PN templates
truncated at the Schwarzschild innermost stable circular orbit (ISCO),
the dashed line corresponds to a search~\cite{lGogginLSC} using
ring-down templates, and the solid line to a search using all three
stages of the binary coalescence using the template bank proposed
here. The horizontal axis reports the total-mass of the binary, while
the vertical axis the distance in Mpc. It is quite evident that, for
binaries with $50 \leq M/M_\odot \leq 140$, the ``coherent search''
using the new template family is considerably more sensitive than any other search
considered here.  In a forthcoming paper \cite{NRDApaper2}, we will
quantify and extend this comparison to include other approaches which
model all the three stages of black hole coalescence such as, for
example the Effective One Body approach \cite{eob1}, or alternatively,
combining the results from separate inspiral and ring down phases
using a coincidence analysis (see e.g.  \cite{sintes-luna}).  It is
worth pointing out that regardless of the sensitivity, a coherent
template bank such as the one proposed here is likely to be
technically easier to implement than a inspiral-ring down coincidence
analysis.  While the method proposed here is just another template
bank search, the inspiral-ring down coincidence would require tuning
the individual inspiral and ring down searches, as well as finding the
appropriate coincidence windows and other pipeline parameters.

However, while this looks promising, we emphasize that it is important
to treat Fig.~\ref{fig:HorDist} as only a preliminary assessment;
fitting factors are not the only consideration for a practical search
strategy. It is also very important to consider issues which arise
when dealing with real data.  For example, false alarms produced by
noise artifacts, might well determine the true sensitivity of the
search, and these artifacts will inevitably be present in real data.
This is however beyond the scope of the present work, and further
investigation is required before we can properly assess the efficacy
of our phenomenological template bank in real-life searches.

\section{Summary and outlook}
\label{sec:summary}

Making use of the recent results from numerical relativity we have
proposed a phenomenological waveform family which can model the
inspiral, merger and ring-down stages of binary black-hole
coalescence. We first constructed a set of hybrid waveforms by
matching the NR waveforms with the analytical PN waveforms. Then, we
analytically constructed phenomenological waveforms which approximated
the hybrid waveforms.  The family of phenomenological waveforms that
we propose was found to have fitting factors larger than 0.99 with the
hybrid waveforms in the detection band of Initial LIGO.  We have also
shown how this phenomenological waveform family can be parametrized
in terms of the physical parameters ($M$ and $\eta$) of the binary, so
that the template bank is, at the end, two dimensional. This
phenomenological waveform family can be used to densely cover the
parameter space, avoiding the computational cost of generating
numerical waveforms at every grid point in the parameter space. We
have also compared the sensitivity of a search using this template
family with other searches. This search might enable us to extend the
mass-range of the present inspiral searches to higher mass ($>
80\,M_\odot$) systems. In the mass-range $50\, M_\odot$ to $140\,
M_\odot$, this search could be significantly more
sensitive than the search using the standard PN inspiral templates and
quasi-normal mode ring-down templates.

The numerical ``look-up tables'' to go from the physical parameters to
the phenomenological parameters (see
Figs.~\ref{fig:BestMatchedAmpParams}
and~\ref{fig:BestMatchedPhaseParams}) can be replaced by analytical
functions of $M$ and $\eta$. This makes it easier to compute the
parameter space metric used for template placement \cite{Owen} and
will be studied in future work.  Our plans for future work also
include the study of the robustness of the matching procedure used to
construct the hybrid waveforms by considering different matching
regions and PN waveforms of different order, and of course, to
eventually construct a realistic search pipeline which incorporates
numerical relativity waveforms in gravitational wave searches.

\label{sec:conc}
\ack 

The authors thank Lisa Goggin and Steve Fairhurst for help in
computing the ring-down horizon distance.  NR computations were
performed with the Belladonna and Peyote clusters of the Albert
Einstein Institute, the Doppler and Kepler clusters of the Jena group,
and at LRZ Munich and HLRS, Stuttgart. This work was supported in part
by DFG grant SFB/Transregio~7 ``Gravitational Wave Astronomy''.  The
Jena group thanks the DEISA Consortium (co-funded by the EU, FP6
project 508830), for support within the DEISA Extreme Computing
Initiative (www.deisa.org).  AMS gratefully acknowledges the support
of the Spanish Ministerio de Educaci\'on y Ciencia research project
FPA-2004-03666 and the Albert Einstein Institute and the University of
Jena for hospitality.  PD thanks the Albert Einstein Institute for
hospitality. The PN waveforms were generated using the LSC Algorithms
Library (LAL), and numerical data-analysis calculations were performed
with the aid of Merlin, Morgane and Zeus clusters of the Albert
Einstein Institute.

\end{document}